%% file: DCM2015Final-Postproceedings.tex
\documentclass[copyright,creativecommons]{eptcs}
\providecommand{\event}{DCM 2015} 
\usepackage{etex}
\usepackage{graphicx}
\usepackage{bussproofs}
\usepackage{proof}
\usepackage{xypic}
\usepackage{pdflscape}
\usepackage[english]{babel}
\usepackage{url}
\usepackage{amssymb,amsmath,amsfonts,txfonts}

\usepackage{tikz}
\usetikzlibrary{snakes,arrows,shapes,automata}

\newtheorem{definition}{Definition}
\newtheorem{theorem}{Theorem}

\newtheorem{example}{Example}
\newtheorem{fact}{Fact}

\def\Set{{\bf SETS}}

\def\LST{{\bf LST}}
\newcommand{\elem}[1]{\ensuremath{\ulcorner #1\urcorner}}


\title{Finiteness and Computation in Toposes}

\author{Edward Hermann Haeusler  
\institute{PUC-Rio \\ Rio de Janeiro, Brasil}
\thanks{The author thanks CAPES and the Petrobras project led by prof. Eduardo Sani Laber at PUC-Rio, for the academic and financial support}
\email{hermann@inf.puc-rio.br}
}

\def\titlerunning{Finiteness and Computation}
\def\authorrunning{E.H.Haeusler}

\begin{document}

\maketitle
\thispagestyle{empty}

\begin{abstract}
Some notions in mathematics can be considered relative. Relative is a term used to denote when the variation in the position of an observer implies variation in properties or measures on the observed object. We know, from Skolem theorem, that there are first-order models where $\mathbb{R}$ is countable and some where it is not. This fact depends on the position of the observer and on the instrument/language the obserevr uses as well, i.e., it depends on whether he/she is inside the model or not and in this particular case the use of first-order logic. 
In this article, we assume  that computation is based on finiteness rather than natural numbers and discuss Turing machines computable morphisms defined on top of the sole notion finiteness. We explore the relativity of finiteness in 
models provided by toposes where the Axiom of Choice (AC) does not hold, since Tarski proved that if  AC holds then all finiteness notions are equivalent. Our toposes do not have natural numbers object ({\bf NNO}) either, since in a topos with a {\bf NNO} these finiteness notions are equivalent to Peano finiteness going back to computation on top of Natural Numbers. The main contribution of this article is to show that although from inside every topos, with the properties previously stated,  the computation model is standard, from outside some of these toposes, unexpected properties on the computation arise, e.g., infinitely long programs, finite computations containing infinitely long ones, infinitely branching computations. We mainly consider Dedekind and Kuratowski notions of finiteness in this article.
\end{abstract}

\section{Introduction}\label{sec:ialc}

Investigations on effectiveness usually follows two non-exclusive
approaches, the model-theoretical and the proof-theoretical one. 
The model-theoretical approach provides a ``model'' $X$, such that, any
(partial) function $F$ from $\mathbb{A}$ into $\mathbb{A}$ is
$X$-effective if and only if there is an instance $X_{F}$ of $X$ that
represents $F$\footnote{In classical theory of recursive functions $\mathbb{A}$ is the set $\mathbb{N}$ of natural numbers.}. The meaning of the ``instance $X_{F}$ that represents $F$'' is
provided informally by stating that for all input $i_X$ submitted to $X_{F}$
produces an output $o_X$, if and only if, $F(i)=o$, for some fixed representation for the input and output data, e.g. numerals or strings simply. The meaning of ``submitting'' and of
``producing'' is also at least informally defined when introducing the
``model'' $X$ main concepts. $X$-effectiveness has to be as close to our intuition on effectiveness as possible. 
The proof-theoretical approach provides a logical theory $T$,  such that,  for any (partial) function $F$ from $\mathbb{A}$ to
$\mathbb{A}$, $F$ is said to be $T$-effective if and only if there is a term $t_{F}$, the codification of $F$,  and a proof that $t_F\in Terms(T)$, such that for every $a\in\mathbb{A}$, $F(a)=b$, if and only if, $t_{F}(a_{F})=b_{F}\in T$. $T$ is presented by a set of axioms and inference rules for deriving
propositions on membership of $T$ and identity between elements of $T$. The main judgments in $T$ are either of the form $t_{F}\in Terms(T)$ or 
$[F\equiv G]\in T$.  

Typical of a model-theoretic approach are Turing machines, while G\"{o}del's (partial) recursive functions follows
the proof-theoretical. Of course, the approaches are not 
purely model-theoretical or proof-theoretical. One can consider
lambda-calculus as a purely proof-theoretical example, by ignoring the
underlying evaluation model that provides the intentional identity relationship. On the other
hand, we can consider the lambda-calculus as a model-theoretical approach if we
focus on the lambda-terms evaluation model. Roger's
theorem stating an  abstract axiomatization for the proof-theoretical approach on
effectiveness  provides stronger\footnote{Stronger than evidences provided by
some concrete models, as those raised since Turing's work.} evidence (read
\cite{Rogers}, page 28, or \cite{Machtey}, isomorphism theorem, chap 4) for Turing-Church thesis: ``A (partial) function from $\mathbb{N}$ to $\mathbb{N}$ is effective, if and only if, it is Turing computable''~\cite{Copeland}. Thus, Roger's theorem can be seen as a meta-theoretical proof-theory based approach for effectiveness.   

This work is based on the observation that finiteness is basic for effectiveness definition. The Turing machine, for example, relies on the finiteness restriction to its tape content, set of symbols and set of states. They have to be finite sets.  ``Programs'', or whatever is used to represent the effective functions,  are finite too, they ``run'' in finite time, they use a finite amount of ``data'', if they are non-deterministic the corresponding non-deterministic range is always finite, etc. Although natural numbers may appear as central in  effectiveness studies, some computational models, do not explicitly mention them, namely  Turing machines. This article also investigates what are the consequences of taking finiteness more basic than natural numbers in this subject.  

In ${\bf ZFC}$\footnote{Zermelo Fraenkel set theory with the axiom of Choice.}, Peano's definition of finite sets uses the set $\mathbb{N}$ of the natural numbers. Nevertheless, definitions of finite in {\bf ZFC}, such as those due to Dedekind or Kuratowski,  do not depend on the  existence of any infinite set, such as $\mathbb{N}$. To the best of our knowledge,  we can say that almost all
model-theoretical approaches for effectiveness are carried out in  {\bf
ZFC}. In 1924, Tarski proved that the many existing and well-known definitions of finite set are equivalent. Tarski mentioned Peano-finiteness, Dedekind-finiteness and some inductive definitions due to M.M. Russell, Sierpinski and Kuratowski, see \cite{TarskiFinite}. A first fact to be noted is  that he had to use the Axiom of Choice in his equivalence proof. 
Another fact is that, due to the duality between finite and infinite, when defining one of these concepts the respective dual is obtained by means of negation. The use of negation adds a logical dimension to this discussion. Thus, besides the Axiom of Choice, the fact that we are inside intuitionistic or classical framework has interesting consequences on the relationship among these mathematical definitions of finiteness. 

Outside the realm of {\bf ZFC}, finiteness is a relative notion.  Relative means that the variation in the position of the observer implies variation in properties or measures on the observed object. This variation may include the instrument used to measure. For example, if we use the language of first-order logic as an analog of a measuring instrument,  Skolem's theorem provides this relativity effect. We have some models where $\mathbb{R}$ is countable, some where it is not. This fact depends on the position of the observer. In this specific case of finiteness it may depend on whether he/she is inside the model or not. Outside {\bf ZFC}, for example, a Dedekind-finite object $O$ can be intuitively infinite, that is,  it can be expressed as $O_1\cup O_2$ with both, $O_1$ and $O_2$ not Dedekind-finite objects, i.e., Dedekind-infinite ones. This is example is shown in this article. One can also argue that first-order logic is not adequate to express Real numbers, this is another point of view. However this kind of discussion may end up into philosophical points that are out of the scope of our approach, as it can be taken in the sequel.  

Based on the relativity of finiteness, we want to start a discussion on what is its real role in Theory of Computation. The methodology is to use a finiteness property as a parameter in Turing-machines definition. For example, by choosing the Dedekind definition of finite we have Turing-Dedekind machines and hence, Turing-Dedekind computable functions. By Tarski's  analysis, these Turing-Dedekind machines cannot be formalized inside {\bf ZFC}, without collapsing into the usual/classical Turing-machines. We will use Category Theory, specifically toposes to obtain these Turing machines definition on top of well-known finiteness definitions, such as Dedekind and Kuratowski. 
Category Theory  ({\bf CT}) is not completely dissociated from set theory
as an alternative theory  for the foundation of mathematics\footnote{The definition of category mention a set/class of objects and a set/class of functions.}. It provides, nevertheless, an
alternative ontology\footnote{Terminology, in philosophical sense.} for
mathematics. In {\bf CT}, classes of objects and morphisms form a category.
Morphisms are typed by domain and co-domain. For example, $A$ and $B$ are
objects in a category ${\mathcal C}$ and $f:A\rightarrow B$ is a morphism in
${\mathcal C}$ having domain $A$ and co-domain $B$. There is a typed
composition ``$\circ$'' operation between  morphisms that has a monoidal
flavor. However, the meta-theory {\bf CT}, apart from the parcel of {\bf ZFC}
that it uses,  does not provide meaning to propositions of the form $A=B$ in
${\mathcal C}$. The meta-theory {\bf CT} only provides meaning to assertions of
the type $f=g$, whenever $f$ and $g$ are morphisms. The whole class of all sets
and all functions between these sets is the archetypal category. Inside this
category (known by \Set) only identity of functions have semantics.
About the objects of \Set, i.e., the sets themselves, it cannot be stated
that any two sets are equal or not equal. The most that can be said is that
they are isomorphic or not\footnote{Let $A$ and $B$ be sets, $A$ is isomorphic to $B$, if and only if, there are
$f:A\rightarrow B$ and $g:B \rightarrow A$, such that, $f\circ g = Id_{B}$ and
$g\circ f=Id_{A}$.}. This changing of perspective is quite interesting since it
provides more ways to compare models of certain concepts formalized on top of
{\bf ZFC} with those formalized on top of {\bf CT}.

We will find out that, at least hypothetically, some Turing-Dedekind machines when observed outside the model, i.e., from the {\bf ZFC} perspective, have infinite set of states and/or infinitely long transition-tables. However, they are finite when observed from inside the model. We could show many other non-standard finite computational models. Because of Tarski's result and the fact that Zermelo-Fraenkel set theory is usually classical, our discussion will be carried in the language {\bf CT}. The use of Local Set Theory and toposes is justified by the fact that the internal logic of a topos is able to easily express set-theoretically based concepts as membership, sub-objects, functions, power-objects. This helps to translate the set-theoretically inspired notions of Dedekind-finiteness and Kuratowksi-finiteness to the Local language. Other categorical approaches to finiteness, as lfp-categories (presented in \cite{lfp-cats} and in \cite{Adamek} in an English comprehensive presentation) lacks internal logic able to define Turing-computable morphisms in the usual set-theoretical analog way. The main contribution of this article is to provide definition of Turing-computable morphism  inside any topos, in a way that finiteness is a parametric notion and does not depend on {\bf NNO}. From this definition we show some examples of toposes, without {\bf NNO} and not satisfying the Axiom of Choice, where from outside them, many unexpected computational properties may hold: 1- Infinitely long Turing machine codes, 2- Infinitely long computations contained in finitely long ones and 3- Infinitely branching computations. In Section~\ref{LST} we briefly explain Local Set Theory and toposes, in Section~\ref{example} we show a Dedekind-finite automaton having an infinite sub-automaton, and how this can be extended to an example of a Dedekind-finite  automaton with arbitrarily many different, and non-isomorphic, infinite sub-automata. We also show, in this topos of automata that there exists an object  $\mathcal{A}$, an automaton,  having only one element $e:1\rightarrow\mathcal{A}$, but $three$ non-empty proper sub-objects. In Section~\ref{finiteness} we state, using the language of Local Set Theory, the finiteness notions of Dedekind-finite, Kuratowski-finite and Peano-finite, as they are quite well-known. In Section~\ref{apendice} we define Turing machines, their computations and the Turing-computable morphisms induced by them, by means of Local Set Theory formulas in an arbitrary topos. In Section~\ref{Squire:sec} we point out the fact that does not seem to exist a natural definition of finiteness, Definition~\ref{Squire}, by showing infinitely many definitions of finite, due to R. Squire. In Section~\ref{conclusion} we remark the existence of some topos where from the outside the above mentioned non-standard computational properties of Turing-machine computations hold. Our assumption on finiteness rather than arithmetic in formalizing the effective is briefly discussed on the light of very relevant articles that discuss the nature of Turing-Church thesis. In Section~\ref{SCT} we remind the reader about the relative inconsistency of a stronger form of Church-Turing thesis with classical logic and the existence of Natural Numbers in the universe of discourse, namely, a topos. Because of this inconsistency we have to either drop out classical logic or the existence of
Natural Numbers object, in order to have such topos. If we define it with a non-classical internal logic we end up revisiting the well studied effective topos, see \cite{Hyland}. On the other hand, if we drop
out {\em Strong Church-Turing thesis}, that is, every morphism in the topos is computable, we obtain the  well studied classical
theory of recursive functions. The last alternatives is to drop out the
existence of a Natural Numbers object and in a naive setting to have finite sets and first-order finite domain logic. From these alternatives, the former, namely, the adoption of a definition of Turing machines by means of only finiteness notions can shed some light on computations expressed 
inside a topos without Natural numbers existence. Section~\ref{apendice} shows how to follow this approach, a contribution of this article.

\section{Local Set Theory and toposes}
\label{LST}
One of the useful aspects of topos theory \cite{Goldblatt,Johnstone} from a logical point of view concerns investigating  the internal logic of toposes, namely categories with some special properties,  by means of localized language, called local set theory (\LST) \cite{Bell}. This has been accomplished by taking any topos as a model of a theory in the language of \LST, which is basically a higher-order typed language. The interpretation of such a theory in the particular topos provide us with a  convenient way of treating the objects of the topos as set-like entities and the morphisms between them as function-like relations between them.

 With the purpose of fixing terminology and provide some (useful) intuition, we write down some definitions. We remind the reader that the definitions in category theory work up to isomorphism.

\begin{definition}[Topos]
A topos is a  category $\mathcal{T}$ having: (1) Terminal Object; (2) Pull-Backs; (3) Exponential Objects; (4) Sub-object Classifier.
\end{definition}

\begin{definition}[Sub-object Classifier]\label{omega}
A sub-object classifier, in a category $\mathcal{T}$, is an object $\Omega$, together with a morphism $\top:1\rightarrow \Omega$, such that, for every  monomorphism $f:B\rightarrow A$, there is a unique morphism $\chi_{f}:A\rightarrow \Omega$, such that, the following diagram is a Pull-Back:
{\small
         \begin{displaymath}
           \xymatrix{
              A\; \ar@{>->}[rr]^{f}\ar[d]^{!} & &  \;B\ar[d]^{\chi_f}  \\
              1\; \ar@{>->}[rr]^{\top} & &\; \Omega 
            }
          \end{displaymath}
}
\end{definition}

The morphism $\top$ plays the rule of the truth-value ``true''. Monomorphisms provide a way of defining sub-objects inside a category. 
Inside a topos, many set-theoretical notions can be categorified, that is, translated to the {\bf CT} language in a way that preserves its original meaning in $\Set$, the category of sets and functions between them. For example, the notion of element of set is in a one-to-one correspondence with functions from a singleton to this set. Each $a\in A$ is associated with the function $f:\{\star\}\rightarrow A$, such that, $f(\star)=a$. The fact that this correspondence can be seen as a bijection between functions from a singleton to $A$ and elements of $A$ allow us to categorify the set-theoretical notion of elements in any category with a terminal object. Terminal objects are the categorical counterpart of singletons.  As a matter of notation, we use (up to isomorphim) $\elem{a}\in A$ to denote that $\elem{a}:1\rightarrow A$ is a morphism in a particular category. The categorification of the empty set is the initial object, since there is one and only one function (in $\Set$) from $\emptyset$ to any other set. As we could expect, categorified notions do not preserve all properties they have in $\Set$, as the two following examples illustrates.

\begin{example}\label{emptyness}
 Initial objects in any category not having a zero object\footnote{A zero object is an object that is initial and terminal.} cannot have elements\footnote{Any morphism from $1$ to $0$ would make them isomorphic.}. Initial objects and empty objects in a category might be the same class. However, in the functor category $\Set^{\cdot\rightarrow\cdot}$, there are objects that are empty and are not initial. An initial object in this category is isomorphic to the only function from $\emptyset$ into $\emptyset$. On the other hand, functions from $\emptyset$ into $A$, for any set $A$, cannot have elements either. The formula $\neg\exists x(x\in X)$ is true if $X$ is assigned to $\emptyset\rightarrow A$ in $\Set^{\cdot\rightarrow\cdot}$.    
\end{example}

\begin{example}\label{extension}
Let $f,g:A\rightarrow B$, two functions in $\Set$, such that $f\neq g$. Consider now the objects $!:\emptyset\rightarrow A$ and $Id_B:B\rightarrow B$ in $\Set^{\cdot\rightarrow\cdot}$ and the commutative diagram below, showing that the  $(!,f)$ and $(!,g)$ are morphisms from $!:\emptyset\rightarrow A$ in $Id_B:B\rightarrow B$ in $\Set^{\cdot\rightarrow\cdot}$.
\begin{displaymath}
                \xymatrix{ A \ar@<0.5ex>[r]^{f}\ar@<-0.5ex>[r]_g  & B \\ 
                           \emptyset \ar[u]_{!}\ar[r]_{!} &  B\ar[u]_{Id_B} }
\end{displaymath}
Because there is no element in $!:\emptyset\rightarrow A$, inside $\Set^{\cdot\rightarrow\cdot}$ it is not possible to falsify the formula $(\forall x\in X(F(x)=G(x)) \rightarrow F=G)$, if $F$,$G$ are assigned to $(!,f)$ and $(!,g)$, respectively, and $X$  to $!:\emptyset\rightarrow A$. Internally $F$ and $G$ are equal, but they are not externally equal.  
\end{example}

Using sub-object classifiers it is possible to locally define equality, the membership relation, existential and the universal quantifier. For any topos they form the semantics of Local Set Theory. The reader can check that Definition~\ref{equality} corresponds to the identity in $\Set$. The other definitions are omitted because of lack of space.  Being a morphism from $A\times A$ into $\Omega$, $=_{A}$ is a predicate. Thus, {\bf \LST}~ has a propositional meaning for $\in_{A}$,  $=_{A}$, $\forall_{A}$ and $\exists_{A}$. They are typed (localized) counterparts of $\in$, $=$, $\forall$ and $\exists$. This is briefly explained in the following subsection.

\subsection{The internal language of a topos}

\begin{definition}[Local Identity]\label{equality}
Consider an  object $A$ in a topos $\mathcal{T}$. Let $\delta:A\rightarrow A\times A$ be the diagonal morphism,  defined as $\langle Id_A,Id_A\rangle$ from the usual categorical cartesian product definition. The sub-object classifier pullback below defines local equality $=_{A}$, in $A$. $=_{A}$ is the characteristic morphism of $\delta$. 
              \begin{displaymath}
                \xymatrix{ A \ar[r]^{\delta_{A}}\ar[d]_{!} & A\times A\ar[d]^{=_{A}} \\ 
                           1 \ar[r]_{\top} &  \Omega }
              \end{displaymath}
\end{definition}

Using the internal logic of the topos, provided by the sub-object classifier, it is possible to define the membership relationship $\in_{A}$, localized in any object $A$ of the topos. This definition strongly relies on the fact that $Hom(A,\Omega)$ represents the collection of sub-objects of $A$\footnote{By the sub-object classifier axiom stated in Definition~\ref{omega} each monomorphism from $A$ to $\Omega$ corresponds to a sub-object of $A$ in a bijective way.}. In fact $Hom(A,\Omega)$ is a Heyting algebra.   

\begin{definition}[Local Membership]\label{membership}
Consider a topos $\mathcal{T}$ and an object $A$ in it. Let $ev_{A}:A\times\Omega^{A}\rightarrow\Omega$ be the evaluation morphism provided by the exponential object $\Omega^{A}$. 
The following instance of the sub-object classifier diagram defines $\in_A$. 
              \begin{displaymath}
                \xymatrix{ \in_A \ar[r]^{inc}\ar[d]_{!} & A\times\Omega^{A}\ar[d]^{ev_A} \\ 
                           1 \ar[r]_{\top} &  \Omega }
              \end{displaymath}
\end{definition}

 So, the local membership to  $A$ is the characteristic morphism of $ev_A$. 

There is a very useful and strong way of building toposes. This is provided by one the fundamental theorem of topos. Its proof can be found in \cite{Goldblatt,Johnstone,Bell}. A category is locally small whenever $Hom(A,B)$ is a set, for any $A$ and $B$ in the category. 

\begin{theorem}
Let $\mathcal{C}$ be a locally small category. $\Set^{\mathcal{C}}$ is a topos.
\end{theorem}

The category $\Set^{\mathcal{C}}$, when $\mathcal{C}$ is a pre-order, is naturally interpreted as sets varying according $\mathcal{C}$. The pre-order works as a temporal structure over each set evolves. When $\mathcal{C}$ is more than a pre-order category, sometimes it is possible to see a kind of topology on any object $A$  induced by the morphism with co-domain $A$. In this case, we have a temporal structure induced by this topology. Anyway, in some cases, $\Set^{\mathcal{C}}$ is naturally equivalent to a category of dynamic systems. Since discrete dynamical systems can be seen as a semantics for computing process, the use of the above functorial category in providing examples for non-standard model of computing is justified. 

\subsection{The logical language related to toposes}

Any topos is (naturally isomorphic to) a model of some local set theory (\LST). In \LST, the notion of type replaces ``set''. In the language of \LST~ each term (including those representing sets) has an associated type. The terminology ``local'' in \LST~ provide us with a scope (locality) to any ``set-theoretical'' operations, such as union, intersection, identity (see Definition~\ref{equality}), membership (see Definition~\ref{membership}), etc. These ``set-theoretical''  operations are only defined for terms of the same type (i.e., locally). Apart from that, the language is very similar to set-theory language, based on the primitive symbols =, $\in$ and the operation $\{\;\mid\;\}$\footnote{The operation $\{\;\mid\;\}$ when applied to a predicate $\phi(x)$ of type $\Omega^{X}$ provides a sub-object $\{x\mid\phi(x)\}$ of $X$.}. The language of \LST~ is defined in the sequel. This presentation follows \cite{Bell} (see pp. 91ff), where the details on how to interpret a local language in an arbitrary topos are provided. 

\begin{definition}[Local Language]
A local language $\mathbf{L}$ is defined by:
\begin{description}
\item[Symbols] the unit symbol 1, the truth-value type symbol $\Omega$, a collection of ground type symbols $A,B,C,\ldots,$ and a collection of function symbols $f,g,h,...$;
\item[Types] the set of types of $\mathbf{L}$ is the least set $\mathbf{T}$ containing 1, $\Omega$, all ground type symbols $A,B,C,\ldots$ and closed under the following operations:
\begin{itemize}
\item For $A\in\mathbf{T}$, the type of power object $PA$ is also in $\mathbf{T}$\footnote{A power object $PA$ internalizes the notion of the ``collection'' of all sub-objects of $A$.};
\item For $A_1,\ldots,A_n\in\mathbf{T}$ , the product type $A_1\times\ldots\times A_n$ is also in $\mathbf{T}$ (for $n=0$, the product type is 1);
\item For $A,B\in\mathbf{T}$, the exponential type $A\rightarrow B$ is also in $\mathbf{T}$.
\end{itemize}
\item[Signatures] Each function symbol $f$ is associated to a signature $A\rightarrow B$, where $A$ and $B$ are types. We use  $f:A\rightarrow B$ to denote this;
\item[Variables] For each type $A$ there is a countable set of variables $V_A$;
\item[Terms] For each type $A$, there is a set $T_A$ of terms of type $A$, defined as follows:
\begin{itemize}
\item $\star\in T_1$;
\item $V_A\subseteq T_A$;
\item For $f:A\rightarrow B$ and $\tau \in T_A$, we have that $f(\tau)\in T_B$;
\item For $\tau_i\in T_{A_i}$ $(i=1,\ldots,n)$ ,we have that $(\tau_1,\ldots,\tau_n)\in T_{A_1\times\ldots\times A_n}$. In the case $n=0$, this term is $\star\in T1$;
\item For $\tau\in T_{A_1\times\ldots\times A_n}$, we have that $\pi_i(\tau)\in T_{A_i}$ $(i=1,\ldots,n)$;
\item For $\varphi\in T_{\Omega}$ and $x\in V_A$, we have that $\{x\mid \varphi\}\in T_{PA}$;
\item For terms $\sigma$ and $\tau$ of type $A$, we have that $\sigma=\tau$ is a term in  $T_{\Omega}$;
\item For terms $\sigma$ and $\tau$ of types $A$ and $PA$, respectively, we have that $\sigma\in\tau$ is a term in $T_{\Omega}$.
\end{itemize}
\end{description}
\end{definition}

Terms of type $\Omega$ are called formulae. We use superscripts to indicate the type of a variable, and, in order to not have overloaded superscripts we allow the omission of some of these superscripts in some terms, since this omission does not interfere in their unique typing. Free and bound occurrences of variables are defined in the usual way. Logical operators can be define as abbreviations, (see \cite{Bell}, p. 70). For example, $\top$ is defined as $\star=\star$; given formulae $\varphi,\psi$ we have that $\varphi\land\psi$ is defined as $(\varphi,\psi)=(\top,\top)$, and $\varphi\Rightarrow\psi$ is defined as $\varphi\land\psi=\varphi$. Quantifiers are also (locally) defined: given a variable $x$ of the appropriate type $A$, 
$\forall x^{A}:\varphi$ is an abbreviation of $\{x\mid\varphi\}=\{x\mid\top\}$. The falsum ($\bot$) is defined as $\forall \omega^{\Omega}:\omega$. Consider a formula $\varphi$, with no occurrence of the variable $\omega$ of type $\Omega$, $\exists x^{A}:\varphi$ is defined as $\forall\omega^{\Omega}:(\forall x^{A}:(\varphi\Rightarrow\omega)\Rightarrow\omega)$. Some terms in a local language represent set-like objects in the Topos.

\begin{definition}[Set-terms]
A set-term is any term of power type $PA$ for some type $A$.
\end{definition} 

Set-theoretical-like definitions are listed in \cite{Bell} (see pp. 83ff). For example: $X\subseteq Y$ is defined as $\forall_{A} x:(x\in X\Rightarrow x\in Y)$; $X\cap Y$ is defined as $\{x\mid x\in X\land x\in Y\}$;
$X\cup Y$ is defined as $\{x\mid x\in X\lor x\in Y\}$, where $X$ and $Y$ are of type $PA$; $A$ is defined as $\{x\mid\top\}$, of type $PA$, with x a variable of type $A$. Thus, for every type symbol $A$, there is a corresponding set-term $A$. The term $\emptyset_A$ is defined as $\{x\mid\bot\}$, of type $PA$, with $x$ a variable of type $A$; $PA$ is defined as $\{x\mid x\subseteq A\}$, of type $PPA$, with $x$ a variable of type $PA$. A more general version of defining set-like objects from set-like objects is provided by the term $\{\tau\mid\varphi\}$, which is defined as $\{x\mid\exists x_1^{A}\ldots\exists x_n^{A}:(x=\tau\land\varphi)\}$. $A\times B$ is defined by $\{(x,y)\mid x^{A}\in A\land y^{B}\in B\}$. $B^{A}$ is defined by $\{f^{P(A\times B)}\mid \forall x^{A}:\exists! y^{B}:(x,y)\in f\}$\footnote{$\exists! y^{A}:\varphi$ is an abbreviation to $(\exists y^{A}:\varphi)\land(\forall x^{A}\forall y^{A}:(\varphi\land\varphi(y/x)\Rightarrow x=y))$, where $\varphi(y/x)$ denotes the replacing of $x$ by $y$ in $\varphi$, usual conditions on replacing free variables applies.}. The type of $B^{A}$ is $PP(A\times B)$. Besides that, to each function symbol $f:A\rightarrow B$ in $\mathbf{L}$ corresponds the set-term $\{(x^{A},f(x))\mid x\in A\}$ of type $B^{A}$. When the topos determines the local language, each  morphism $f:A\rightarrow B$ is associated to a set-term $f$ of type $B^{A}$, allowing us to represent morphisms as functions-like entities, in terms of ``sets'' of ordered pairs. Finally,  infinite versions of these operations, for indexed families, are such that, $\oplus_{i\in I} X_{i}$ is defined as $\{(i^{I},x^{X_i})\mid i\in I\land x\in X_i\}$. In \cite{Bell} it is shown a deductive system, in sequent-style, to derive (draw conclusion) formulas from set of formulas. It is shown how any consistent\footnote{A set of formulas is consistent, if and only if, it does not derive $\bot$.} set of formulas in \LST~ gives raise to a syntactical topos (in the style of a Herbrand term model), and how from syntactical models one can derive a completeness theorem for \LST~ logical consequence.  

A notion of validity of a formula in a topos is defined in the expected way. Consider a mapping $M$ from a Local Language $\mathbf{L}$ into a topos $\mathcal{T}$, such that, the types $\mathbf{T}$ are mapped into the objects of $\mathcal{T}$, $\Omega$ is mapped into a sub-object classifier of $\mathcal{T}$ ($\Omega_{\mathcal{T}}$), $1$ is mapped into a terminal object of $\mathcal{T}$, products are mapped into products, function symbols are adequately mapped into morphisms,  and variables of type $X$ into morphisms from $M(1)$ into $M(X)$. This mapping is recursively extended to a mapping $\hat{M}$ from terms into objects and morphisms of $\mathcal{T}$. 
Thus, for each set-like term $t$, of type $PA$, $\hat{M}(t)$ is mapped to a corresponding sub-object of $\hat{M}(A)$. Terms of type $\Omega$ are mapped to morphisms from $M(1)$ to $M(\Omega)$, that are truth-values inside the topos. $M$ can be seen as denotational interpretation of $\mathbf{L}$ into $\mathcal{T}$. Thus, having denotations for $t_1$ and $t_2$ of type $A$, $M(t_1)$ and $M(t_2)$ sub-objects of $M(A)$, the denotation of the formula $t_1=t_2$ can be considered as true, if and only if, the monomorphism related to $M(t_1)$ is equal to the one related to $M(t_2)$, since Category Theory provides equality between morphism, this is unproblematic at first sight. However, the (local) equality  can be defined inside a topos. As it was shown in Definition~\ref{equality}, the denotation of $t_1=t_2$ is itself a general element of (a morphism from $M(1)$ to $M(\Omega)$), and hence a morphism from $1_{\mathcal{T}}$ into $\Omega_{\mathcal{T}}$, a truth value itself. Since the equality defined in Definition~\ref{equality} is a local notion, it is hardly the case that the truth value $M(t_1=t_2)$ is the same of $M(t_1)=M(t_2)$. Example~\ref{extension} confirm this and Example~\ref{emptyness} provide a typical case of a categorification of a set-theoretical concept that does not have the same truth value on every topos. 
The concept of monomorphism, used here and not defined yet, is of a different kind. Monomorphisms are categorifications of injective functions: $f:A\rightarrow B$ is a monomorphism (mono), if and only if, for every pair of morphisms $h,g:C\rightarrow A$, such that $f\circ h=f\circ g$, then $h=g$. In a topos monomorphisms and injections coincide. In a general category this is not the case. In a topos every morphism that is injective and surjective is an isomorphism. Of course, this is the case in $\Set$, since $\Set$ is a topos. It is clear that there are properties expressed in \LST~ that hold in some topos and does not in $\Set$.

\section{Motivating Examples}
\label{example}
M-Sets form an universe that can be identified with the class of  automata and morphisms between them.
Let $M$ be a monoid and  $A$ be a set. 
$\mathcal{A}:A\times M\rightarrow A$ is a M-Set, if and only if, $\mathcal{A}(a,m\star n) = \mathcal{A}(\mathcal{A}(a,m),n)$, where $\star$ is the monoid operation.
Let $\mathcal{A}:A\times M\rightarrow A$ and $\mathcal{B}:B\times M\rightarrow B$ be $M$-Sets. A function $F:A\rightarrow B$, such that,  $F(\mathcal{A}(x,m))=\mathcal{B}(F(x),m)$ is a morphism in $Sets^{M}$. We denote it by $\mathcal{F}:\mathcal{A}\rightarrow\mathcal{B}$.
An M-Set (monoid actions) can be seen as a family $(\mathcal{A}_{\sigma})_{\sigma\in M}$ of functions from $A$ into $A$. Note that we use $A$ for both, the family of functions and their domain and co-domain sets. The functions must satisfy: $A_{m\star n}(x)=A_{m}(A_{n}(x))$.
 $F:\mathcal{A}\rightarrow\mathcal{B}$ is  morphism, if and only if, $F(\mathcal{A}_m(x))=\mathcal{B}_m(F(x))$. $M$-Sets and $M$-Sets morphisms form a category that is a topos.
Let $\mathcal{A}$ be the finite automaton at the right side of Figure~\ref{auto-um}(a). It  can be uniquely defined by the actions $m_a$, $m_b$ and $m_c$ on the set $\{q_1,q_2,q_3,q_4\}$, such that:
$\mathcal{A}_a(q_1)  =  q_2, 
\mathcal{A}_a(q_2)= q_4,
\mathcal{A}_a(q_3)  = q_4,
\mathcal{A}_a(q_4)  =  q_4,
\mathcal{A}_b(q_1)  =  q_2,
\mathcal{A}_b(q_2)  =  q_3,
\mathcal{A}_b(q_3)  =  q_3,
\mathcal{A}_b(q_4)  =  q_4,
\mathcal{A}_c(q_1)  =  q_3,
\mathcal{A}_c(q_2)  =  q_2,
\mathcal{A}_c(q_3)  =  q_2,
\mathcal{A}_c(q_4)  =  q_4$. It can be proved that every $\Sigma^{\star}$-Set is an automaton on $\Sigma^{\star}$, not necessarily finite. 
\begin{center}
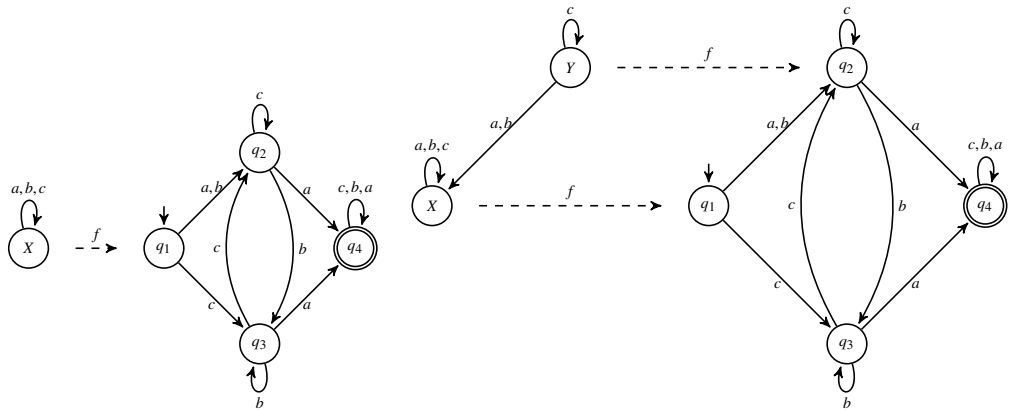
\begin{figure}[h]
{\tiny
\begin{tabular}{cc}
\begin{tikzpicture}[->,>=stealth',shorten >=1pt,auto,node distance=1.8cm,semithick]
  \tikzstyle{every initial by arrow}=[initial text=,->]
  \node[initial above,state] (q1) {$q_1$};
  \node[state] (X) [left of=q1] {$X$};
  \node[state] (q2) [above right of=q1] {$q_2$};
  \node[state] (q3) [below right of=q1] {$q_3$};
  \node[state,accepting] (q4) [above right of=q3] {$q_4$};

  \path[dashed, shorten >=10pt, shorten <=10pt] (X) edge node [above] {$f$} (q1);
  \path (q1) edge node [above] {$a,b$} (q2)
        (X) edge [loop above] node {$a,b,c$} (X)
  
        (q1) edge node [below] {$c$} (q3)
        (q2) edge [bend left] node {$b$} (q3)
        (q3) edge [bend left] node {$c$} (q2)
        (q2) edge [loop above] node {$c$} (q2)
        (q3) edge [loop below] node {$b$} (q3)
        (q2) edge node[above] {$a$} (q4)
        (q3) edge node[below] {$a$} (q4)
        (q4) edge [loop above] node {$c,b,a$} (q4);
\end{tikzpicture}
& 
\begin{tikzpicture}[->,>=stealth',shorten >=1pt,auto,node distance=2.6cm,semithick]
  \tikzstyle{every initial by arrow}=[initial text=,->]
  \node[initial above,state] (q1) {$q_1$};
  \node[state] (Y) [above left of=q1] {$Y$};
  \node[state] (X) [below left of=Y] {$X$};
  \node[state] (q2) [above right of=q1] {$q_2$};
  \node[state] (q3) [below right of=q1] {$q_3$};
  \node[state,accepting] (q4) [above right of=q3] {$q_4$};

  \path[dashed, shorten >=10pt, shorten <=10pt] (Y) edge node [above] {$f$} (q2);
  \path[dashed, shorten >=10pt, shorten <=10pt] (X) edge node [above] {$f$} (q1);
  \path (q1) edge node [above] {$a,b$} (q2)
        (Y) edge node [above] {$a,b$} (X)
        (X) edge [loop above] node {$a,b,c$} (X)
        (q1) edge node [below] {$c$} (q3)
        (q2) edge [bend left] node {$b$} (q3)
        (q3) edge [bend left] node {$c$} (q2)
        (q2) edge [loop above] node {$c$} (q2)
        (Y) edge [loop above] node {$c$} (Y)
        (q3) edge [loop below] node {$b$} (q3)
        (q2) edge node[above] {$a$} (q4)
        (q3) edge node[below] {$a$} (q4)
        (q4) edge [loop above] node {$c,b,a$} (q4);
\end{tikzpicture}
\end{tabular}
\caption{Monomorphisms: Left (a) from one-state automaton $\mathcal{B}$ into $\mathcal{A}$. Right, (b) two-state $\mathcal{B}$ into $\mathcal{A}$}
\label{auto-um}}
\end{figure}
\end{center}

 What are the  automata $\mathcal{B}$ that can be related to $\mathcal{A}$ by monomorphisms.  Consider $\mathcal{B}$ with only one state $X$. 
A mono $F:\mathcal{B}\rightarrow\mathcal{A}$ is a function $f:\{X\}\rightarrow\{q_1,q_2,q_3,q_4\}$ that is equivariant. 
$f(X)=q_4$ is the only possibility, for $f(B_{\sigma}(X))=A_{\sigma}(f(X))$, see Figure~\ref{auto-um}(a). In Figure~\ref{auto-um}(b), $\mathcal{B}$ has two states $X$ and $Y$. A mono $F:\mathcal{B}\rightarrow\mathcal{A}$ is a function $f:\{X,Y\}\rightarrow\{q_1,q_2,q_3,q_4\}$ equivariant and injective.
With $f(X)=q_4$, there is no way to have $f(B_{\sigma}(Y))=A_{\sigma}(f(Y))$, for any choice, such that $f(Y)= \{q_3,q_1,q_2\}$.

\begin{center}
\begin{figure}
{\tiny
\begin{tikzpicture}[->,>=stealth',shorten >=1pt,auto,node distance=2.6cm,semithick]
  \tikzstyle{every initial by arrow}=[initial text=,->]
  \node[initial above,state] (q1) {$q_1$};
  \node[state] (X) [left of=q1] {$X$};
  \node[state] (Y) [above left of=X] {$Y$};
  \node[state] (Z) [below left of=X] {$Z$};
  \node[state] (q2) [above right of=q1] {$q_2$};
  \node[state] (q3) [below right of=q1] {$q_3$};
  \node[state,accepting] (q4) [above right of=q3] {$q_4$};

  \path[dashed, shorten >=10pt, shorten <=10pt] (Y) edge node [above] {$f_2$} (q2);
  \path[dashed, shorten >=10pt, shorten <=10pt] (X) edge node [above] {$f_2$} (q1);
  \path[dashed, shorten >=10pt, shorten <=10pt] (Z) edge node [above] {$f_2$} (q3);
  \path (q1) edge node [above] {$a,b$} (q2)
        (Y) edge node [above] {$a$} (X)
        (Z) edge node [above] {$a$} (X)
        (Z) edge [loop below] node {$b$} (Z)
        (X) edge [loop above] node {$a,b,c$} (X)
        (Z) edge [bend left] node {$c$} (Y)
        (Y) edge [bend left] node {$b$} (Z)
        (q1) edge node [below] {$c$} (q3)
        (q2) edge [bend left] node {$b$} (q3)
        (q3) edge [bend left] node {$c$} (q2)
        (q2) edge [loop above] node {$c$} (q2)
        (Y) edge [loop above] node {$c$} (Y)
        (q3) edge [loop below] node {$b$} (q3)
        (q2) edge node[above] {$a$} (q4)
        (q3) edge node[below] {$a$} (q4)
        (q4) edge [loop above] node {$c,b,a$} (q4);
\end{tikzpicture}
\caption{Monomorphism from a three-state automaton $\mathcal{B}$ into $\mathcal{A}$}
\label{figura1depoisrevisao}
}
\end{figure}
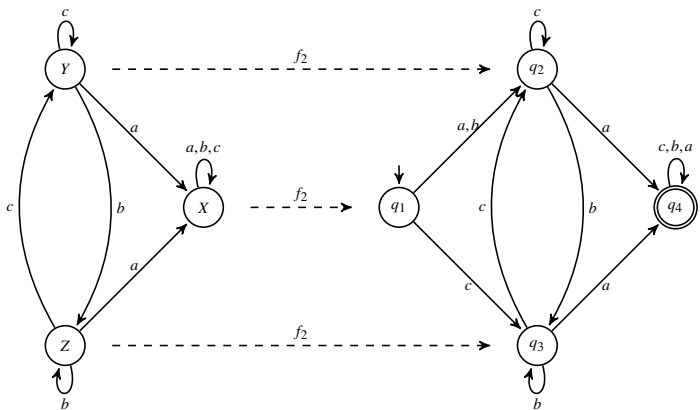
\end{center}
There are only two monomorphisms $F_1,F_2:\mathcal{B}\rightarrow\mathcal{A}$ that are equivariant  from an automaton with 3 states.
They correspond to the functions $f_1$ and $f_2$: $f_1(X)=q_4$ and $f_2(X) = q_4$
and $f_1(Y)=q_3$ with $f_2(Y) = q_2$, and, 
$f_1(Z)=q_2$  with $f_2(Z) = q_3$ . See Figure~\ref{figura1depoisrevisao}.

The $\{a,b,c\}^{\star}-Set$ $\mathcal{A}$ has only one element $1:\rightarrow\mathcal{A}$, but $three$ non-empty proper sub-objects. In {\bf ZFC} this statement cannot be true. Any set with only one element cannot have a power-set with three elements. This is yet another manifestation of the relativity of the concept of finiteness. 
Let us show an example on Dedekind notion of finiteness.
In \LST, we express that $f$ is an isomorphism by $Iso(f^{B^A})\Leftrightarrow \exists h((f\circ h =_{B^B}Id_{B}\wedge (h\circ f=_{A^A}Id_{A}$. We express that $f$ is a monomorphism by $Mono(f^{B^A})\Leftrightarrow \forall h^{A^C}\forall g^{A^C}((f\circ h =_{B^C}f\circ g)\rightarrow h=_{A^C}g)$.
In a topos $\mathcal{T}$, $A$ is D-finite in $\mathcal{T}$, if and only if, $\forall f\in A^{A} (Mono(f)\Rightarrow Iso(f))$ holds in $\mathcal{T}$.


\begin{example}
Let $M$ be the free monoid generated by $\{m_i/\mbox{$i\in\mathbb{N}$}\}$, $A$=$\{a_n/n\in\mathbb{N}\}$, $B$=$\{b_n/n\in\mathbb{N}\}$, and $C=A\cup B$.
The automaton $\mathcal{C}:M\times C\rightarrow C$ is the following. 
{\tiny
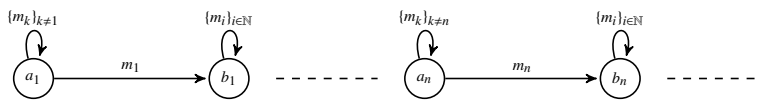
\begin{center}
\begin{tikzpicture}[->,>=stealth',shorten >=1pt,auto,node distance=2.6cm,semithick]
  \tikzstyle{every initial by arrow}=[initial text=,->]
  \node[state] (a1) {$a_1$};
  \node[state] (b1) [right of=a1] {$b_1$};
  \node[state] (an) [right of=b1] {$a_n$};
  \node[state] (bn) [right of=an] {$b_n$};
  \node[state, draw=none] (fim) [right of=bn] {};

  \path (a1) edge node [above] {$m_1$} (b1)
        (a1) edge [loop above] node {$\{m_k\}_{k\neq 1}$} (a1)
        (b1) edge [loop above] node {$\{m_i\}_{i\in\mathbb{N}}$} (b1)
        (an) edge node [above] {$m_n$} (bn)
        (an) edge [loop above] node {$\{m_k\}_{k\neq n}$} (an)
        (bn) edge [loop above] node {$\{m_i\}_{i\in\mathbb{N}}$} (bn);
  \draw[-,dashed, shorten >=10pt, shorten <=10pt] (b1) edge node [-] {} (an);
  \draw[-,dashed, shorten >=10pt, shorten <=10pt] (bn) edge node [-] {} (fim);
\end{tikzpicture}
\end{center}
}
\end{example}
We show that any injective  morphism $\mathcal{F}:\mathcal{C}\rightarrow\mathcal{C}$ is the identity $Id_{\mathcal{C}}$ (see Example~\ref{infiniteAutomaton}).
Thus, $C$ is Dedekind-finite, no matter the set-theoretical cardinality of $A$ and $B$. If  $B$ is an infinite set then there is 
$G:B\mapsto B$, an injective function that is not bijective. Thus, $\mathcal{B}(m_n,b_k)=G(b_k)$ is not Dedekind-finite\footnote{Remember that a M-Set is an action $M\times B\mapsto B$.}. That is, there is a Dedekind-finite ($\mathcal{C}$) object that has a D-infinite sub-object ($\mathcal{B}$). This can be extended for any $n\in\mathbb{N}$, such  that, $\mathcal{C} = \mathcal{A}\cup\mathcal{A}_1\cup\mathcal{A}_2\cup\ldots\cup\mathcal{A}_n$, $\mathcal{C}$ is Dedekind-finite and $\mathcal{B}_i$, $i=1,\ldots,n$, are D-infinite, see Example~\ref{infiniteinfiniteautomaton}.

\section{Finiteness in Local Set Theories}
\label{finiteness}
The main discussion in this article, is the role played by finiteness in computability. We are sure that this notion is essential when defining any computation model, even it is used implicitly. From the literature on topos theory, we are aware of the fact that the well-known notions of finite, namely, Dedekind-finite, Kuratowski-finite and Peano-finite, for example are not equivalent, see \cite{Kock1975}. This in fact means that some of these notions correspond to non-finite (infinite) extensions, externally, in some toposes. Besides that, such notions have counter-intuitive properties as we will mention in the sequel. 

As a starting point, we write down, in \LST, sometimes with the help of diagrams these finiteness notions. 

The first item of the above list,  Dedekind finiteness or D-finiteness as defined in Section~\ref{example},  is quite interesting. There we show D-finite objects that are externally infinite. Besides that, a result due to Johnstone~\cite{Johnstone:AlgUniv} shows that any sub-object classifier is D-finite. However, there are many toposes that have infinitely many truth-values. Sheaves over a topological space $S$ have Opens(S) as the ``set'' of truth-values, for example. 
One can argue that D-finiteness is obtained by negating D-infinity, and hence, it is strongly dependent on whether the topos is classical or not. Kuratowski-finiteness, in contrast, is based on a positive aspect of finiteness. Intuitively, an object is Kuratowski-finite, if and only if, we can provide an inductive proof that it is finite. This induction is based on the facts that the empty object and the singletons are finite and any binary union of finite objects is finite too. 

\begin{definition}[Kuratowski-finite]
In a topos $\mathcal{T}$, $A$ is Kuratowksi-finite (K-finite), if and only if, the following holds: 
{\small
\[
\forall z\in \Omega^{\Omega^A}([0\rightarrow A]\in z\wedge\forall a\in A(\{a\}\in z)\wedge 
      \forall y\in \Omega^{A}\forall y^{\prime}\in\Omega^{A}((y\in z\wedge y^{\prime}\in z)\Rightarrow (y\cup y^{\prime})\in z)\Rightarrow [id_{A}]\in z)
\]
}
\end{definition}

Considering $K(A)$ as the sub-object of $\Omega^{A}$ formed by the K-finite subobjects  of $A$, we have (cf. \cite{Johnstone}):
a) 0 e 1 are K-finite; b) If $f:A\rightarrow B$ is an epimorphism and $A$ is  K-finite then  $B$ is  K-finite too; c) $B,C\in K(A)$, if and only if, $B\cup C\in K(A)$;
d) If $B$ and $C$ are  K-finite then   $B+C$ and $B\times C$ are  K-finite; e) If $A$ is K-finite and $B\in \Omega^A$ and $B$ is  complemented then  $B$ is  K-finite too; f) $X$ is  K-finite, if and only if, $K(X)$ is K-finite.

In both finiteness notions above it is possible to have a non-finite sub-object of a finite  object. It happens in any topos that does not have always complemented sub-objects. This is the case in all non-classical toposes.  Example of these toposes are used in the computational definitions in the sequel and in the examples in Section~\ref{example}. The third notion of finiteness is reported here only as matter of completeness. It involves the existence of a {\bf NNO} in the topos. We are not considering that {\bf NNO} is an essential starting point for defining a computational model. In a topos having {\bf NNO}, any Peano-finite object is a K-finite and a D-finite object.
A natural number object is a rather categorical definition of natural numbers in a category. It states that there is an object $N$ with two morphisms $0:1\rightarrow N$ and $s:N\rightarrow N$, such that, for any other object $A$ and morphisms $g:1\rightarrow A$ and $h:A\rightarrow A$, there is one and only one morphism $f:N\rightarrow A$, such that the primitive recursive equations $f(0)=g(0)$ and $f(s(n))=h(f(n))$ hold categorically. {\bf NNO} allows us to define any primitive recursive function inside its category. Thus, in a topos having a {\bf NNO}, the relation less-than is defined, by primitive recursion,  stating that $\langle$ is a sub-object of $N\times N$. We define $[p]\approx \{0,\ldots,p-1\}$, as a sub-object of $N$ using $\Omega$ axiom:
              \begin{displaymath}
                \xymatrix{ [p] \ar[rrr]^{!}\ar[d] & & & 1 \ar[d]^{p} \\ 
                           \langle \ar[r] & N\times N \ar[r]^{\pi_2} & N\ar[r]^{suc} & N }
              \end{displaymath}

\begin{definition}[Cardinal-finite]
In a topos $\mathcal{T}$ with {\bf NNO}, $A$ is c-finite, or Peano-finite,  if and only if, $A$ is isomorphic to $[p]$ for some  $p:1\rightarrow N$.
\end{definition}

\section{Strong Church-Turing thesis inside toposes}
\label{SCT}
In this section we discuss the relationship among {\bf NNO} existence, classical logic reasoning and the (strong)
 Church-Turing thesis inside an arbitrary topos. The intention of this discussion is to reinforce conceptual and philosophical 
aspects on the Church-Turing thesis  drawn in \cite{Sieg} that point out to the direction we follow in this article, when we consider finiteness
more basic than Natural Numbers in the study of effectiveness.

The following argument is found in \cite{Shapiro:11} and
specifically in \cite{Mclarty} using the language of {\bf
  CT}. Consider the {\em Strong Church Thesis} ({\bf SCT})
``Every function from $\mathbb{N}$ in $\mathbb{N}$ is computable.''
A function is computable if and only
if there is a program that computes it. Any program can be
expressed by its code that, in its turn can be viewed as a natural
number. Thus, {\bf STC} is expressed by the following number-theoretical
formula:
\[\forall f \, \exists p\,\forall n \, \exists y \cdot 
(T(p,n,y) \land Out(y)=f(n)),\] 
where $T(p,n,y)$ is Kleene's $T$ predicate and $Out(y)$ is Kleene's
output function. The meaning of $T(p,n,y)$ is that $p$ when running over $n$ produces the Turing machine configuration $y$. Using the Peano Arithmetic we can obtain
$T$ and $Out$ as primitive recursive predicate and function
respectively. Thus, in any topos with a Natural Number object, $T$ and
$Out$ are primitive recursive too. In Local Set Theory,
{\bf SCT} is of form: \[\forall f^{\mathbb{N}^\mathbb{N}}\exists
p^{\mathbb{N}}\forall n^{\mathbb{N}}\exists y^{\mathbb{N}}(T(p,n,y)
\land Out(y)=_{\mathbb{N}}f(n)).\]

Considering an arbitrary topos having a Natural Numbers object, and having classical logic as internal
logic, it can be shown that {\bf SCT} is inconsistent with the statements above, namely, in this topos {\bf SCT} cannot hold. Using the fact
that classical logic satisfies the law of excluded middle, the
definition for $g$ as follows: \[ g(n) = \left\{
\begin{array}{ll} m+1 & \mbox{if~~ $\exists j^{\mathbb{N}}(T(n,n,j) \land
Out(j)=m)$} \\ 0 & \mbox{otherwise} \end{array} \right.  \]    is provable to
be defined for every $n$. Thus $g$ is a total function, and hence by {\bf SCT}
has a program $p$. However, any program $p\in\mathbb{N}$ that implements $g$ is
such that $g(p) = (g(p)) + 1$, since there is $j$, such that, $T(p,p,j)$ and
$Out(j)=m$ and $(g(p))=m+1$. This is not possible, so {\bf SCT} is inconsistent
with the law of the excluded middle in a topos having a Natural Numbers object. 

Let us analyze the alternatives when defining a topos of computable morphisms.  It may have the following properties:
\begin{enumerate}
\item\label{um} The internal logic of the topos is classical;
\item\label{dois} Every morphism in the topos is effective, i.e., {\bf SCT} holds in the topos;
\item\label{tres} The topos has a Natural Numbers object.
\end{enumerate}
We have just shown that~\ref{dois} is inconsistent with~\ref{um}
and~\ref{tres}, so either we drop out classical logic or the existence of
Natural Numbers object. If we define a topos with a non-classical internal logic we end up revisiting the well studied effective topos, see \cite{Hyland}. On the other hand, if we drop
out item~\ref{dois}, we obtain with~\ref{um} and~\ref{tres} the also well studied classical
theory of recursive functions. The last alternative is to drop out the
existence of a Natural Numbers object. In a naive setting, \ref{um}
and~\ref{dois} together entail finite sets and first-order finite domain logic.
The definition of Turing machines by means of only finiteness notions can shed some light on computations expressed 
inside a topos without {\bf NNO}. This is what the following section does. 
 Finally, the definition of Turing machines by means of only finiteness notions can shed some light on computations expressed 
inside a topos without Natural numbers.

\section{Turing Machines in Local Language}\label{apendice}

In this section we detail our finiteness-parametric Turing-computable functions inside toposes that may not have natural numbers objects. The potential cases of internal standard Turing machines representing the external non-standard are: (1) Infinitely long programs; (2) Infinitely many branching, and; (3) Infinitely long traces inside finitely long ones; as remarked in Section~\ref{conclusion}. In the sequel we sometimes use T.M. for denoting a  Turing machine.  

In this subsection we express Turing Machines by means of a local language. We consider that finiteness is essential for the definition of any computational model. Instead of fixing a specific finiteness notion, we use it as a parameter. So, consider $fin(X)$ a predicate that defines a finiteness notion. 
In the usual (set-theoretical) definition of a Turing Machine, as found in \cite{Hopcroft}\footnote{In \cite{Hopcroft} the alphabet distinguishes input and working symbols, here this distinction is not relevant} for example,  we have 
$\langle Q,\Sigma,\{q_o\},\{q_f\},\delta\rangle$, and,   $\delta\subseteq 2^{\Sigma\times Q\times \{\leftarrow,\rightarrow\}\times \Sigma\times Q}$. $Q$ and $\Sigma$ have to be finite and non-empty sets of states and symbols,  and $\delta$, the transition function,  is finite as a consequence of the finiteness of $Q$ and $\Sigma$, and the fact that the power set of a finite set is finite. Besides that, the elapsed time of any meaningful computation has to be finite. The behavior of the Turing Machine is described by the definition of $\hat{\delta}$\footnote{$\hat{\delta}$ is the function from T.M. configurations into the power set of T.M. configurations that represents the transitive/reflexive closure of the transition $\delta$. It is usually defined in Theory of Computation textbooks by means of $\vdash^{\star}$ the relation between T.M. configurations entailing the transitive closure of the one-step transition table, see \cite{Hopcroft} for example}. We proceed in defining the semantics of a Turing Machine, in a topos  without {\bf NNO}. It is important to remember that in a topos with {\bf NNO}, the finiteness notions collapse to Peano-finiteness.
\begin{definition}[Turing Machine in \LST]
Let  $\Sigma$ and $Q$ be types and  $X^{\Sigma}$, $Y^{Q}$, $q_o^{Q}$ and $q_f^{Q}$ variables, then $T$, a variable of type $Q\times\Sigma\times Q\times Q\times[Q\times\Sigma\rightarrow P(Q\times\Sigma\times 2)]$ is a Turing Machine, if and only if, $\pi_1(T)=X\land \pi_2(T)=Y\land(\pi_3(T)\in X)\land(\pi_4(T)\in X)\land fin(X)\land fin(Y)$. We denote this predicate as $TM(T)$. 
\end{definition}

The type $2$, in the above definition,  is used to denote $Right$ and $Left$, the directions of a possible moving in a one-step transition of the Turing Machine. The definition above regards to non-deterministic Turing machines. As we will see in the sequel, there are non-standard external behavior in case we consider deterministic (only) Turing machines too. The possibility of hypercomputational external behavior happen in both cases, deterministic and non-deterministic, but the sort of hypercomputational behavior caused by non-determinism does not seem to be emulated by the deterministic version. 

Note that we do not have the type of the Turing Machines in this formalization. We have chosen this way, in order to not have to fix a type for states and alphabets. Whenever we refer to a Turing Machine T, it is a variable $x$ of the type $Q\times\Sigma\times Q\times Q\times[Q\times\Sigma\rightarrow P(Q\times\Sigma\times 2)]$, for some types $Q$ and $\Sigma$, such that $TM(x)$. In the sequel, we show, in an informal way,  the definition of the behavior of a Turing Machine $T=\langle X,Y,q_o,q_f,Z\rangle$, with adequate typing corresponding to $Q\times\Sigma\times Q\times Q\times[Q\times\Sigma\rightarrow P(Q\times\Sigma\times 2)]$. 

Consider a type $Pos$, a variable $Z^{Pos}$ and a variable $X^{\Sigma}$. $H^{\Sigma^{Pos}}$ is a tape, if it satisfies $Tape(H^{\Sigma^{Pos}})\Leftrightarrow fin(\{p^{Pos}\mid H(p)\neq \mbox{``\;''}\})\land\neg fin(Pos)$\footnote{The symbol ``\;'' denotes blank in the cell.}. In a Turing machine tape, the non-blank cells must be only finitely many. Besides that, there is no limit in the tape for storing symbols.  

\begin{definition}[Closure]\label{closure}
The behavior of T is $\hat{\delta}$ (a variable of type defined in the sequel). It is specified by means of the following variables (implicitly locally quantified) and predicates. At the end we have the morphism computed by T. 
\begin{itemize}
\item $H:Pos\longrightarrow \Sigma$, such that $Tape(H)$
\item $f_{\delta}:\mathcal{P}(Q\times Pos\times T_{fin} )\longrightarrow \mathcal{P}(Q\times Pos_{fin}\times Q)$, such that $f_{\delta}(S_1)=S_2\Leftrightarrow$ \linebreak $S_2=\{s\mid \exists s_1\in S_1(\pi_4(T)(\pi_1(s_1),\pi_2(s_1),\pi_3(s_1))=s\land Tape(\pi_3(s_1))\}$, where  $\pi_4(T)(X,Z,H)$ is an usual translation for the local typed language of the one-step of a Turing Machine, possibly non-deterministic,  $T$, in state $X$, on position $Z$, and tape $H$. 
\item $Config(X)=\{S\subseteq X:\mbox{$f_{\delta}(S)\subseteq S$}\}$
\item $i:St(X)\hookrightarrow \mathcal{P}(X)$ has left adjoint $\mathcal{O}:\mathcal{P}(X)\rightarrow St(X)$, $\mathcal{O}(Z)=\bigcap\{ S\in St(X):\mbox{$Z\subseteq S$}\}$
\item $\mathcal{O}\circ i$ is a closure operator, thus we have $\hat{\delta}=\mathcal{O}\circ i$
\item $f=\{\langle x^{\Sigma^{\star}},y^{\Sigma^{\star}}\rangle / \mbox{$\langle q,z^{Pos},y^{T}\rangle\in \hat{\delta}\circ \langle q_0^{Q},o^{Pos},x^{T}\rangle$ and $final(q)$}\}$. Here, to obtain $\Sigma^{\star}$, the type of strings, a similar closure definition and application has to be done. 
\end{itemize}
\end{definition}
This definition uses a  technique from~\cite{Orbits} to denote orbits in arbitrary toposes, without {\bf NNO}. It is important to note that if the topos has a {\bf NNO}, then the closure defined above is just Kleene closure.

Having defined the concept of a morphism being Turing computable in a topos, we have that:
In $Sets^{G}$, $G$ a particular free group,  with $fin(A)=Dedekind(A)$, then 
$Sets^{G}\models\exists W(fin(W)\land\neg fin(\mathcal{P}(W))$. Hence in $Sets^{G}$, a T.M. with states in $W$ is a non-standard computational model possibly with a  D-infinitely long program, or a D-infinitely branching non-deterministic behavior. 

In $Sets^{0\rightarrow 1}$ with $fin(A)=Kuratowski(A)$, then 
$Sets^{0\rightarrow 1}\models \exists W(fin(W)\land \exists V(V\subseteq W\land \neg fin(V)))$.
 Thus, a T.M. with states in $W$ can be  a non-standard computational model able to K-finitely compute on K-infinitely long transitions.

We provided here arguments in favor of a kind of non-standard finiteness phenomena inside $M$-Sets. They justify in details what is observed in the conclusion of our work (Section~\ref{conclusion}).

\begin{fact}\label{infiniteAutomaton} 
Let $M$ be the free monoid generated by $\{m_i/\mbox{$i\in\mathbb{N}$}\}$
$A$=$\{a_n/n\in\mathbb{N}\}$, $B$=$\{b_n/n\in\mathbb{N}\}$, and $C=A\cup B$
Let $\mathcal{C}:M\times C\rightarrow C$ be the action of $M$ on $C$, such that, 
\[
\mathcal{C}(m_n,x)=\left\{\begin{array}{ll}
                         a_k & \mbox{if $x=a_k$ and $k\neq n$} \\
                         b_k & \mbox{if $x=a_k$ and $k=n$} \\
                         b_k & \mbox{if $x=b_k$}
                         \end{array}\right.
\]
Any injective morphism $\mathcal{F}:\mathcal{C}\rightarrow\mathcal{C}$ is the identity $Id_{\mathcal{C}}$. This is justified by observing that, if $\mathcal{F}$ was not the identity, then there would exists $n$, such that, $f(a_n)\neq a_n$. Thus, 
$f(a_n)=x\neq a_n$ $\Rightarrow$ $\mathcal{C}(m_n,x)=x$, for $a_n$ is the only element of $A\cup B$ changed by the action $C$\footnote{Note that we confuse the set with the action.} then, $f(b_n)=f(C(m_n,a_n))=C(m_n,f(a_n))=C(m_n,x)=x=f(a_n)$. This cannot be possible. 
\end{fact}

Let $G:B\mapsto B$ be an injective function that is not bijective. The action $\mathcal{B}(m_n,b_k)=G(b_k)$ is not Dedekind-finite. Thus, $\mathcal{B}$ is a Dedekind-infinite object inside a Dedekind-finite object in $M$-Sets.

\begin{example}\label{infiniteinfiniteautomaton}
It is also possible to have arbitrarily many  disjoint infinite sub-objects of $C$
Let $\mathcal{C}:M\times (A\cup B\cup D)\rightarrow (A\cup B\cup D)$ be the action of $M$ on $C$, such that, 
\[
\mathcal{C}(m_n,x)=\left\{\begin{array}{ll}
                         a_k & \mbox{if $x=a_k$ and $k\neq n$} \\
                         b_n & \mbox{if $x=a_k$ and $k=n$} \\
                         b_k & \mbox{if $x=b_k$ and $k\neq n$} \\
                         c_n & \mbox{if $x=b_k$ and $k=n$} \\
                         c_k & \mbox{if $x=c_k$}
                         \end{array}\right.
\] 
Any injective morphism $\mathcal{F}:\mathcal{C}\rightarrow\mathcal{C}$ is the identity $Id_{\mathcal{C}}$. The same argument used in fact~\ref{infiniteAutomaton} is used to prove that if 
$f(a_n)=x\neq a_n$ $\Rightarrow$ $C(m_n,x)=x$, then, $f(b_n)=f(C(m_n,a_n))=C(m_n,f(a_n))=C(m_n,x)=x=f(a_n)$. That is not possible. 
On the other hand if each of $B$ and $D$ and $A$ are infinite sets then there are $G:B\mapsto B$ and $H:D\mapsto D$ 
 injective functions that are not bijective. The actions $\mathcal{B}(m_n,b_k)=G(b_k)$ and $\mathcal{D}(m_n,c_k)=H(c_k)$ prove that both, $\mathcal{B}$ and $\mathcal{D}$ are not Dedekind-finite.
\end{example}

In \cite{Eilenberg:70}, a categorical presentation of recursiveness is provided
using {\bf CT}. It axiomatizes categories able to define primitive recursive
morphisms in a completely abstract way. Using the internal language of the
category it is possible to precisely define any primitive recursive function.
This work is very interesting, since, it joins in a quite harmonious way a
model-theoretic definition with a proof-theoretic one. The identity present in
the meta-theory provides meaning for a theory of equality between intentionally
distinct ways of defining the primitive recursive functions. Besides that no
mention on a concrete numerical system of even richer  definition of natural
number is needed, but primitive recursiveness.

\section{There are infinitely many finiteness definitions}
\label{Squire:sec}
As a matter of curiosity, we show that there are infinitely many definitions of finite. The following definition is from \cite{Squire1997}.

\begin{definition}[Squire-finite]\label{Squire}
Let $p\in\mathbb{N}$, $p\neq 0$. Let $\phi_p$ be any formula expressing that there is at most $p$ things, for example, $\phi_p=\exists x_1\exists x_2\ldots\exists x_p\forall y(y=x_1\lor y=x_2\ldots\lor y=x_p)$, or, $\phi_p=\bigvee_{i<j\leq p+1} (x_i=x_j)$. Thus, $A$ is $L_{p}$-finite, if and only if: 
\begin{itemize} 
\item $L_p(A)$ is the upper-semi-lattice generated by $\{S:\mbox{$S\in 2^A$ and $S\models\phi_p$}\}$
\item $\hat{id_{A}}:1\rightarrow \Omega^{A}$ factors as:
\[
\xymatrix{
1 \ar@{>>}[dr] \ar@{>}[rr]^{\hat{id_{A}}} &   &  \Omega^{A} \\
&  L_p(A) \ar@{>->}[ur] & }
\]
\end{itemize}
\end{definition}
Note that the formula $\bigvee_{i=1,p} (y=x_i)$ can be also taken as $\phi_p$. The quantifiers are not essential to this end, since each $x_i$ and $y$ has to be assigned to an ``element'' $e:1\rightarrow A$ of $A$, anyway.

The first thing that we have to observe is that in {\bf Sets} every set that is $L_p$ finite is finite. $L_1$-finiteness is basically Kuratowski-finiteness. Thus, any finite set is $L_1$-finite and hence it is also $L_p$-finite, for any $1\leq p$. On the other hand, if a set $A$ is $L_p$-finite, this means that $A\in L_p(A)$, since the arrow $\hat{id_{A}}$ factors through $L_p(A)$ says this. Besides this, as we already said,  Tarski proved that Dedekind, Kuratowski and Peano-finiteness are equivalent. Thus, as $A\in L_p(A)$ means that the cardinality, $card(A)$ is less than or equal to $2^{2^p}$ in sets, then $A$ is Peano-finite and hence finite in a very intuitive sense. 

\begin{fact} Let $A$ be a set that it is $L_p$-finite, for some $p\neq 0$. So $A$ is Peano-finite, Dedekind-finite and Kuratowski-finite. 
\end{fact}   

Besides that, we have the following facts.

\begin{fact} Let $q\leq p$. If $A$ is $L_q$ finite then it is $L_p$ finite. 
\end{fact}

\begin{fact} In $M$-Sets, there are objects $L_p$ finite that are not $L_q$ finite, $q\leq p$.
 \end{fact}

The first fact is obvious from the definition of $\phi_p$ and the observation that $L_p(A)$ includes $L_q(A)$, thus every $A$ that factors through $L_q(A)$ factors through $L_q(A)$ too. The second fact is a bit involving. Consider the monoid $M=\langle\{1\leq \ldots\leq p\}, \lor, 1\rangle$. Thus, $i\lor j=max(i,j)=j\lor i$. In the category of the $M-Sets$, $M$ itself is an $M-Set$. The sub-objects of $M$ as an $M-Set$, are in one-to-one correspondence with the subsets of $\{1,\ldots,p\}$, closed by the operation $\lor$. This sub-objects are the final segments of the form $\{k,\ldots,p\}$, $k\leq p$. The lattice is a linear order, thus, the upper-semi lattice generated by the sub-objects are themselves. In this sense, a $L_q$-finite sub-object of $M$ is formed by the unions of the generators $\{k,\ldots p\}$, $k\leq p$. Then, 
this $L_q$-finite set is not $L_{q-1}-finite$, for the generators this turn are $\{2,\ldots,q\}, \{3,\ldots,q\}, \ldots, \{q\}$. They do not include $1$, so $\{1,\ldots,q-1\}$ does not factor through $L_{q-1}$. Hence, this $M$ is not $L_{q-1}$-finite.   

\section{Conclusion}
\label{conclusion}
 In some mathematical universes, e.g. M-Sets, finite objects may not share essential properties with their {\bf Sets} counterparts. Finiteness is relative.   
The first example in Section~\ref{example} was carried out using a rather intuitive notion of finiteness. Given a property $F(S)$ that provides a finiteness definition for $S$, in {\bf \LST}, and a topos $\mathcal{T}$, we can express  Turing machines (TM) and their semantics in $\mathcal{T}$ itself. In Definition~\ref{closure} some effort is need to define TM-semantics without natural numbers. $f:A\rightarrow B$ in $\mathcal{T}$ is TM-computable, if and only if, there is a TM that has $f$ as semantics. What is interesting in this general definition of TM computability is that, without appealing to infinitely long computations and programs, as well non-deterministic branching, for the TMs are internally finite, programs and branching can be externally infinite. We intend to see how this is related to computation on infinite-time Turing machines (\cite{infinite}), but at the present stage of the research and the lack of space we can not provide any definite result.

Finally, we would like to comment on some recent analysis on the status of Turing-Church thesis, by Sieg,  Dershowitz and Gurevich, in~\cite{Sieg} and~\cite{Dershowitz:08}, for example.

Wilfried Sieg, in~\cite{Sieg}, observes that Turing-machines and $\lambda$-calculus are not explicitly number-theoretic based models of computation, instead, they take seriously boundedness and locality as basic concepts in computation. Boundedness and locality are finiteness aspects of the computational underlying model. This is justified in~\cite{Sieg}, where Church's citations on the adequateness of the restrictions of finiteness on the machines regarded the ordinary notion of computing (see pg. 6 of~\cite{Sieg}). This is precisely the main restrictions Turing states in his seminal article. The argument that Church follows, according~\cite{Sieg}, is to reserve Turing's analytic steps: ``a human calculator, provided with pencil and paper and explicit instructions, can be regarded as a kind of Turing machine''. It seems to be clear, at least for Sieg, Dershowitz and Gurevih, that the unique aspect of Turing's work on analytically exploiting the limitations of the essential  abilities of the human computing agent, in order to take basic principles for his machine definition, point out to finiteness restrictions (boundedness and locality) on  computors (human computing agent) and computing machines. The work of Sieg is important when it points out that finiteness is a quite relevant aspect of computing. We cannot forget, however,  that the approaches by G\"{o}del and Kleene have taken  the study of effectiveness to the number-theoretic level. At this level, finiteness was granted for free. Under this number-theoretic approach, our work is meaningless, since a category with {\bf NNO} and effectiveness defined on top of Natural Numbers, is not different from the classical recursion theory. A topos with {\bf NNO}, as seen in Section~\ref{SCT}, does not provide much alternative model-theoretic properties when compared to classical recursion theory. We can say that we followed the foundational/historical  arguments that are in~\cite{Sieg} and~\cite{Dershowitz:08} to an extent that they do not consider. Both works turn back to the number-theoretic aspect of computing in order to either follow an empirically based discussion (\cite{Sieg}) or to provide a kind of proof of Church-Turing thesis (\cite{Dershowitz:08}). We can observe that taking Natural Numbers as the basic elements when defining computational models, was an historical moving that led to many interesting and important equivalences. However, what does it remain when we drop Natural Numbers ?  This is the kind of question that we started to answer by modeling computation inside toposes as we shown here.

\section{Acknowledgements}

We would like to thank prof. Luiz Carlos Pereira and the project {\em Logic in Ilha Grande} that motivates the initial ideas on the research reported here. We 
are thankful to Peter Arndt, whose seminars pointed out to me the many facets of finiteness provided by topos-theoretic models. We are also grateful to prof Gilles Dowek for the discussions on the Turing-Dedekind machines and the viability of their  use on standard computational models. The statement of what is  {\em relative concept} in mathematics was firstly provided by him in an informal meeting. Bruno Lopes, Hugo Macedo, Ranieri Batista and Jefferson Santos have read previous versions of this article. We thank them very much. Finally we would like to thank the referees of the DCM2015 first submitted version, particularly the referee that pointed out other works that consider seriously the concept of finiteness as being more basic than the Natural Numbers in effectiveness definition.

\nocite{*}
\bibliographystyle{eptcs}
\bibliography{DCM2015-Toposes}

\end{document}